\documentclass[twocolumn,prl,showpacs ]{revtex4}

\usepackage{amsmath}
\usepackage{amssymb}
\usepackage{graphicx}
\usepackage{wrapfig}
\usepackage{color}

\newcommand{\refeq}[1]{Eq. \ref{#1}}
\newcommand{\sigm}{{\sigma}}

\newcommand{\mem}{{\mathrm{mm}}}
\newcommand{\bE}{{\bf E}}
\newcommand{\bT}{{\bf T}}
\newcommand{\bn}{{\bf n}}

\newcommand{\bv}{{\bf v}}

\newcommand{\bff}{{\bf f}}

\newcommand{\im}{{\mathrm i}}

\begin{document}

\title{Lipid membrane instability and poration driven by capacitive charging}
\author{Jonathan T. Schwalbe$^1$, Petia M. Vlahovska$^2$ and Michael J. Miksis$^1$}
\affiliation{$^1$Department of Engineering Sciences and Applied Mathematics, Northwestern University, Evanston, IL 60202, USA\\
$^2$Thayer School of Engineering, Dartmouth College, Hanover, NH 03755, USA}

\date{\today}

\begin{abstract}
A new model for the interaction of an electric pulse with a lipid membrane is proposed. Using this model we show that when a DC electric pulse is applied to an insulating  lipid membrane separating fluids with different conductivities, the capacitive charging  current through the membrane drives electrohydrodynamic flow that destabilizes the membrane. The instability is transient and decays as the membrane charges. 
The bulk conductivity mismatch plays an essential role in this instability because it results in a different rate of charge accumulation on the membrane's physical surfaces. Shearing stresses created by the electric field acting on its own induced free  charge are non-zero as long as the charge imbalance exists.  Accordingly, the most unstable mode is related to the ratio of membrane charging time and the electrohydrodynamic time.

% 87.16.D-,87.16.dj,  87.16.-b}

\end{abstract}

\maketitle

Cells and cellular organelles are enveloped by lipid membranes, whose integrity is crucial for the cell viability. External electric fields can cause membrane breakdown and cell death. However, a  controlled application of electric pulses can induce transient pores in the cell membrane, which reseal after the pulse is turned off. This reversible electroporation is of great interest for biomedical technologies such as gene transfection because the pores enable the delivery of exogenous molecules (drugs, proteins,  DNA) into living cells \cite{Neumann-Sowers-Jordan:1989}.  However, controlled electroporation remains elusive  because the physical mechanisms underlying membrane response to electric fields are poorly understood \cite{Weaver:1996, Teissie:2005, softmatter:2009}. Here we propose a new model for the interaction of an electric pulse with a lipid membrane. The model sheds new light on the electroporation process and is applicable to a broad range of biological interfaces.

The lipid bilayer membrane is impermeable to ions and  acts as a capacitor when an electric field is applied.
Charges carried by conduction accumulate at the physical surfaces and   a potential difference across the membrane builds up. In the case of a non-conducting membrane subjected to a uniform DC electric field, the transmembrane potential  increases as
\begin{equation}
\label{Vm}
\begin{split}
V_m=V\left[1-\exp\left(-t/t_m\right)\right]
 \end{split}
\end{equation}
where $t_m$ is a characteristic charging time for the capacitor and $V$, in the case of a planar membrane, is the applied potential difference. It is generally accepted that the membrane porates when the the transmembrane potential exceeds a critical value of about  one Volt\cite{Needham-Hochmuth:1989}. In a typical electroporation experiment \cite{Riske-Dimova:2005}, $t_m\sim 1ms$. This is comparable and in certain cases even longer than the duration of the  electric pulse, yet poration may occur at voltages lower than the critical one. 
This observation has been attributed to the presence of an initial  tension in the membrane. A possible alternate explanation is that the membrane becomes unstable even before the critical voltage is reached. In this Letter we propose a new mechanism for such  an instability. We examine a non-conducting membrane and show that  upon application of a uniform DC electric field a transient  induced charge  appears on the membrane.
The electric field acting on this charge gives rise to membrane in-plane electric stresses, which move the lipids and adjacent fluids.
The resulting electrohydrodynamic flow   can enhance a small perturbation of the membrane shape. This mechanism is similar to the destabilization of a fluid interface due to a finite time for charge relaxation \cite{Melcher-Smith:1969}.

Although the  theoretical modeling of the dynamics of  lipid membranes in steady electric fields has received considerable attention \cite{Sens-Isambert:2002,  Ambjornsson:2007, Vlahovska-Dimova:2009, Lacoste:2009}, we are not aware  of any study that has examined the effect of the transient electric field on the stability of the membrane. Our model is fundamentally different from earlier work, in which various mechanisms for electroporation  have been examined: pore nucleation \cite{DeBruin-Krassowska:1999a, Krassowska:2007}, membrane thinning \cite{Crowley:1973, Dimitrov:1984,  Isambert:1998}, 
negative tension  \cite{Sens-Isambert:2002} 
or induced charge electro-osmotic flow \cite{Lacoste:2007, Lacoste:2009, Lacoste:2010}. Moreover, unlike recent  electromechanical models \cite{ Isambert:1998, Sens-Isambert:2002, Lacoste:2007, Lacoste:2010}, which have considered a conducting membrane, 
 we analyze an {{insulating}} membrane.

\begin{figure}[h]
\centerline{\includegraphics[width=3in]{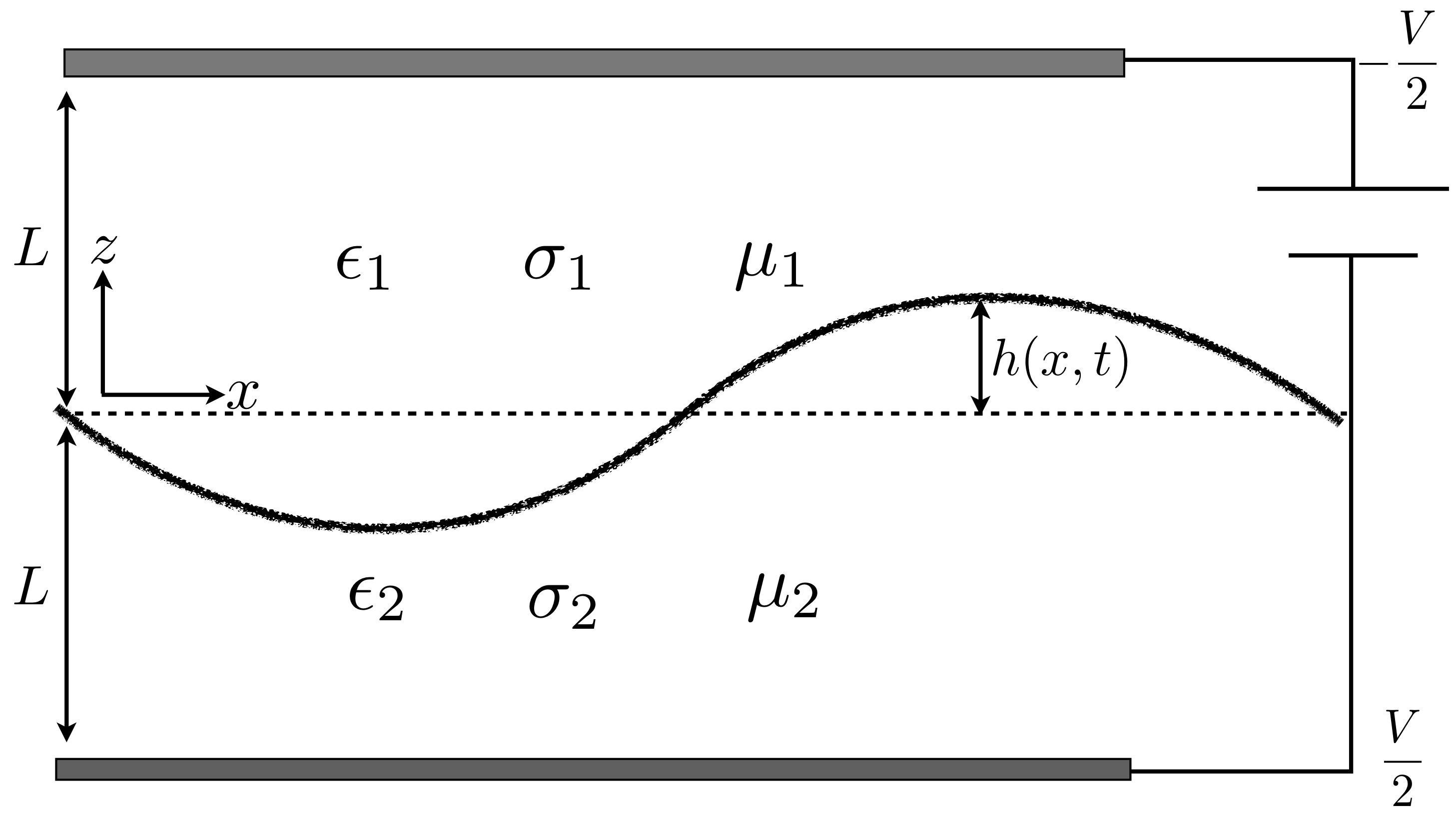}}
     \caption{\footnotesize  A sketch of the problem. }
      \label{fig1}
\end{figure}

Let us consider a planar non-conducting membrane formed  by a charge-free lipid bilayer with  dielectric constant $\epsilon_\mem$. The bilayer thickness is about $d\sim~5nm$, thus on macroscopic length scales the membrane can be regarded as a two-dimensional surface with capacitance $C_m=\epsilon_\mem/d$. The membrane separates two fluids: a superphase of viscosity $\mu_1$, conductivity $\sigm_1$, and dielectric constant $\epsilon_1$, and a subphase characterized respectively by $\mu_2$,  $\sigm_2$, and  $\epsilon_2$. The membrane is subjected to a perpendicular electric pulse with magnitude $E_0=V/2L$, where $V$ is the applied potential, and $2L$ is the distance between the electrodes. The problem is sketched in Figure \ref{fig1}. 

Upon application of the electric field, bulk phases become electroneutral on a very fast time scale given by $t_{c,k}=\epsilon_k/\sigm_k$ (where $k=1,2$ denotes the top or bottom fluids).
Accordingly, the electric potential $\phi$ is a solution of Laplace's equation and the equations of bulk fluid motion have no electric terms, i.e.  the electromechanical coupling occurs only at the interface. This is in essence the leaky dielectric model developed by G. I. Taylor \cite{Saville:1997}.  Note that even though electrokinetic transport  is not explicitly included, the effect of space charge is accounted for in an aggregate sense\cite{Baygents-Saville:1988, Zholkovskij:2002}. 

The electrostatic problem is specified by  $\nabla\cdot \bE_k=0$ and $\nabla\times \bE_k=0$, which is equivalent to $\bE_k=-\nabla\phi_k$ and $\nabla^2\phi_k=0$ with boundary conditions at the electrodes $\phi_1(L)=-V/2$, $\phi_2(-L)=V/2$. At the membrane interface, $z=h(x, t)$, conservation of normal currents requires 
\begin{equation}
\label{bc:current}
\begin{split}
C_m\frac{dV_m}{dt}=&\sigm_1 \bn\cdot\bE_1=\sigm_2 \bn\cdot \bE_2\,,
 \end{split}
\end{equation}
where  Ohmic conduction is assumed for the bulk fluids.   $ V_m=\phi_2-\phi_1 $ is the transmembrane potential and $\bn$ is the unit normal vector. At $t=0$ when the pulse is applied the potential is continuous, $V_m(t=0)=0$. 
The difference in bulk fluid conductivities gives rise to a transient imbalance in the charge densities on the two sides of the membrane (since charges are brought at a different rate)
\begin{equation}
\label{charge}
Q=\bn\cdot (\epsilon_1\nabla\phi_1-\epsilon_2 \nabla\phi_2). 
\end{equation}
$Q$ is not to be mistaken with the charge of the capacitor; if the capacitor is fully charged $Q=0$. 

The  electric field stresses the interface, which responds by deformation. The fluid  flow accompanying the interface motion is described by the Stokes equations, because at the length scales of the membrane undulations viscosity dampens fluid acceleration. The velocity, $\bv_k$, and the pressure, $p_k$, fields satisfy
\begin{equation}
  \nabla p_k = \mu_k\nabla^2 \bv_k,\quad
  \nabla\cdot\bv_k = 0\,.
  \label{eq:floweqs}
\end{equation}
 At the electrodes the fluid satisfies the no slip boundary condition and
the velocity is continuous across the interface. The interface moves with the velocity of the surrounding fluid,
\begin{equation}
\label{kin cond}
	\frac{\partial h}{\partial t} = \bv_s\cdot\bn\left(1+\partial_xh^2\right)^{1/2}\,,\quad \bv_1=\bv_2\equiv\bv_s\,.
\end{equation}
The surface flow is area-incompressible, $\nabla_s\cdot\bv_s=0$. 

The coupling of the electric and hydrodynamics fields at the interface is reflected by the stress balance  
\begin{equation}
\label{stressb}
(p_2-p_1)\bn +\bn\cdot\left[\left[\bT^{hd}+\bT^{el}\right]\right]=\bff^{m}\,,
\end{equation}
where $\left[\left[...\right]\right]$ denotes a difference between fluid 1 and fluid 2. The Maxwell stress tensor is  $T^{el}_{ij}=\epsilon\left(E_iE_j-1/2 E^2\delta_{ij}\right)$, $T^{hd}_{ij}=\mu\left(\partial_iv_j+\partial_jv_i\right)$ is the viscous stress tensor, and $\bff^{m}$ are the membrane tractions. In the simplest model (Helfrich-Canham)
\begin{equation}
\label{memtrac}
\bff^{m}=\left[-\kappa\left(2H^3-2HK+\nabla_s^2 H\right)+\Sigma H\right]\bn+\nabla_s \Sigma \,,
\end{equation}
where $\kappa$ is the bending rigidity, $H$ and $K$ are the mean and Gaussian curvatures. Here $\Sigma$ is the membrane tension, a Lagrange multiplier which enforces the surface-incompressibility \cite{Seifert:1999}.

Next we proceed to analyze the linear stability of a planar membrane subjected to an electric pulse. 
The evolution of a membrane perturbation in the $x$-direction with a wavenumber $q$ depends on the interplay of several physical process: the decrease of the effective interfacial charge (which occurs on a characteristic time scale corresponding to the charging of the membrane capacitor, $t_m$),  shape distortion (which occurs on electrohydrodynamic time scale, $t_{ehd}$),  and curvature relaxation on a time scale $t_\kappa$
\begin{equation}
t_m=C_mL/\sigma_2 \,,\quad  t_{ehd}={\mu_2}/{(\epsilon_2 E_0^2)}\,,\quad t_\kappa = \mu _2/(\kappa q^3).
\end{equation}
In a typical electroporation experiment, the fluids are aqueous salt solutions, $\sigm\sim 10^{-4} S/m$,  $C_m\sim 0.01 F/m^2$, and $E_0\sim 1 kV/cm$. Hence, $t_m \sim t_{ehd}\sim  1ms$ while $t_c\sim1\mu s$, i.e.  both the electrohydrodynamic and the capacitor charging times are much longer that the bulk charge relaxation time, which justifies the quasi-static approximation.

The base state of the problem is a flat interface, $h=0$, stressed by a perpendicular electric field; any variable $g$ is  expanded in a series of the form  $g= g^{(0)}(z,t) +g^{(1)}(z,x,t)+...$, where  $g^{(1)}=\sum_q g_q(z,t) \exp(\im qx)$. 
The problem solution involves the following steps. 
First, we solve the electrostatic problem to find the electric tractions exerted on the membrane. Second, we solve for the fluid flow needed to satisfy the stress balance, \refeq{stressb}. Finally, the interface evolution is obtained from the kinematic condition, \refeq{kin cond}. For the membrane to deform, the normal electric stress  must be nonvanishing $T_{zz}^{(1)}=E^{(0)} E^{(1)}$. This requires that the base state electric potential is non-uniform, which occurs only if there is current through the membrane; for a purely capacitive interface this is possible only during the period of charging.

Henceforth, all variables are nondimensionalized. The characteristic time scale is the capacitor charging time $t_m$, the length scale is $L\equiv V/E_0$, and stresses are scaled by the viscous fluid stress $\mu_2/t_m$.
We define the conductivity ratio $R=\sigm_1/\sigm_2$, permittivity ratio $S=\epsilon_1/\epsilon_2$, and viscosity ratio $\lambda=\mu_1/\mu_2$.

{\it{Base state: Flat membrane }}

The solution for the transient electric field about a flat membrane yields  
\begin{eqnarray}\label{flat}
\begin{split}
&\textstyle{\phi_1^{(0)}=-\frac{1}{2}+(z - 1)E_1^{(0)}(t)}\,,\\
&\textstyle{\phi_2^{(0)}=\frac{1}{2}+(z+1)R E_1^{(0)}(t) }
\end{split}\end{eqnarray}
where $E_1^{(0)}(t)=e^{-\alpha t}/(1+R)$ and $\alpha= R/(1+R)$. Hence, the effective surface charge and transmembrane potential are
\begin{equation}
Q^{(0)} =(S-R)E_1^{(0)}(t)\,,\quad V_m^{(0)} = 1 - e^{-\alpha t}.
\end{equation} 
Initially the potential is continuous across the membrane, but the interface is  charged (positively for $R/S<1$, i.e., if the bottom fluid is more conducting, and negatively for $R/S>1$).  At times much longer than the capacitor time $t_m$ the membrane becomes fully charged, the electric field in the bulk vanishes and the transmembrane potential becomes equal to the applied potential.

{\it{Fluctuating membrane: leading order analysis  of the electric field effect on membrane undulations}}

The solution of Laplace's equation for the electric potential in a bounded domain gives
\begin{eqnarray}\begin{split}
	\phi_{1,q}^{(1)} &= A_q(t)\sinh[q(z-1)] \,,\\
	\phi_{2,q}^{(1)} &= RA_q(t)\sinh[q(z+1)]\,.
\end{split}\end{eqnarray}
The coefficient $A_q(t)$ is determined from the  conservation of current condition (\refeq{bc:current}), which at this order couples to the amplitude of the interface perturbation $h_q(t)$
\begin{equation}\label{bc:perturbedcurrent}
	\frac{d A_q}{d t}  + \frac{qR\coth(q)}{1+R}A_q = \frac{1}{\sinh(q)}\frac{1-R}{\left(1+R\right)}\frac{\partial}{\partial t}(h_q E_1^{(0)}).
\end{equation}
Note that the surface incompressibility requires $v_x(z=0)=0$.

At this order in perturbation, the kinematic condition is  $dh/dt=v_z^{(1)}(z=0)$. Inserting the hydrodynamic and electric stresses in  the normal stress balance (\refeq{stressb} and \refeq{memtrac}) yields
\begin{eqnarray}\begin{split}
\label{bc:perturbedstress}
2(1+\lambda)\frac{d h_q}{d t} =\frac{1+2q^2-\cosh 2q}{2q+\sinh 2q}\Big(\Pi_1\left(q^3+\zeta q\right)h_q\\
-\Pi_2(R^2-S)A_q(t) E_1^{(0)}(t) \cosh q\Big)
\end{split}\end{eqnarray}
where $\Pi_1=t_m/t_\kappa$ ($t_\kappa$ is defined based on $L$), $\Pi_2=t_m/t_{ehd}$, and $\zeta=L^2 \Sigma_{eq}/\kappa$.  We see that while the first term on the right hand side describes relaxation, the second term is destabilizing if $E_1^{(0)}(t)\neq 0$ and there is a mismatch in the conductivities of the fluids.
The tangential stress balance  gives a nonuniform membrane tension $\Sigma_q(t)=(R^2-S) E_1^{(0)}(t)[A_q(t)+E_1^{(0)}(t) h_q(t)]$
which influences the dynamics at next order. 
 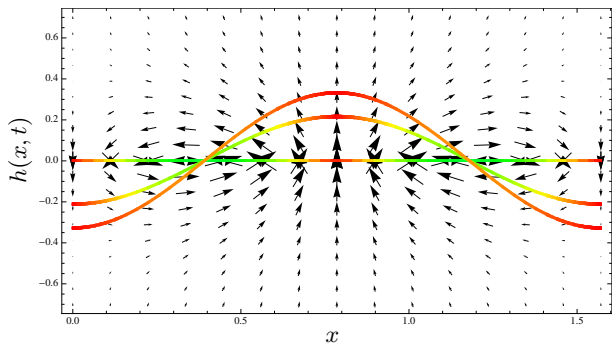
\begin{figure}[h]
\centerline{\includegraphics[width=3in]{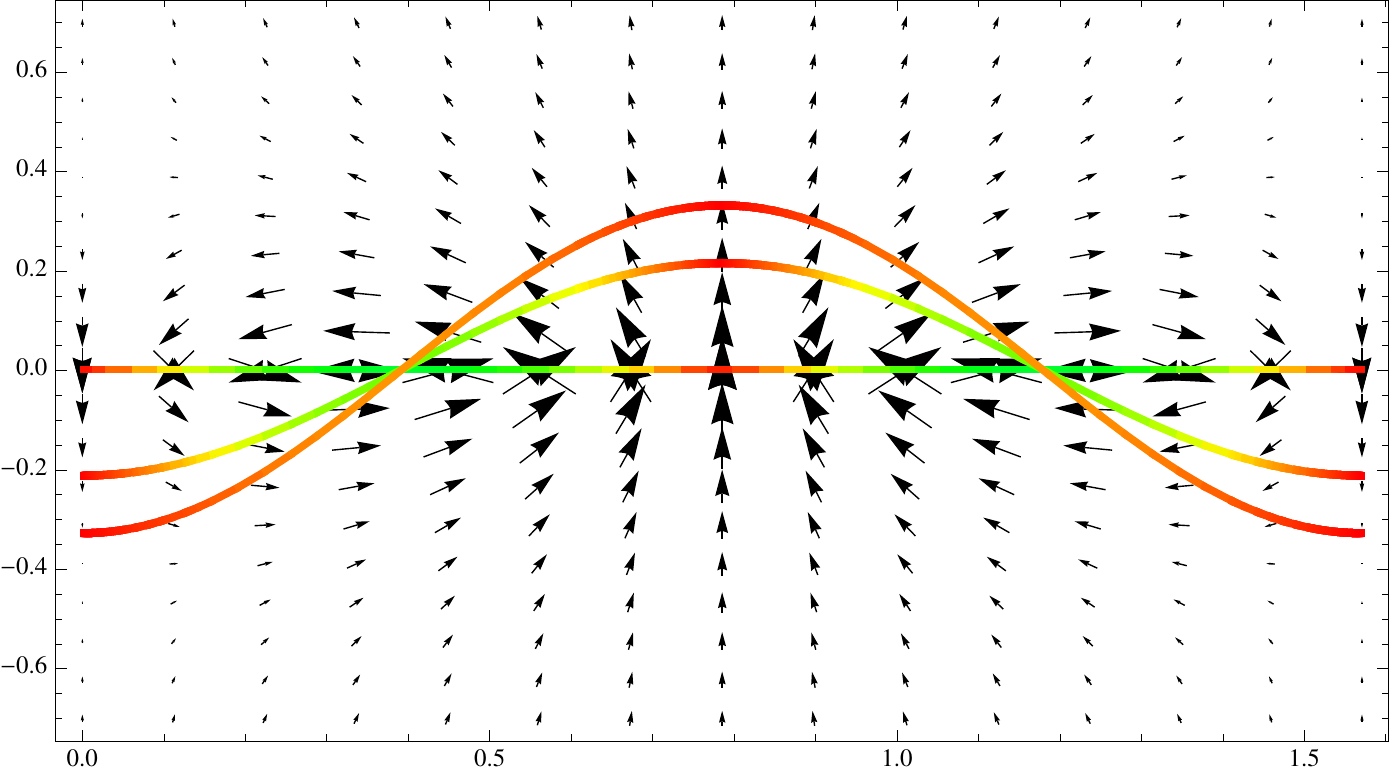}}
\begin{picture}(0,0)(0,0)
\put(0,5){$x$}
\put(-120,65){\rotatebox{90}{$h(x,t)$}}
\end{picture}
      \caption{\footnotesize  (Color online) Increasing shape undulation in response to a  perturbation in the transmembrane potential. The arrows show the electric field at $t=0$.  The color coding reflects the magnitude of electric shearing stresses (red = zero shear stress). Interface profiles at $t=0$, $t=0.5$, and $t=2$.  Parameters as in Figure \ref{fig4}.}
      \label{fig3}
\end{figure}
To illustrate this time dependent behavior,  \refeq{bc:perturbedcurrent} and \refeq{bc:perturbedstress} are solved numerically  for $q=4$ and an initially flat interface (initial condition small $A_4\ne 0$ and $h_4=0$).  In response to the electric pulse the membrane deforms; this is accompanied with variations of the effective interfacial charge and shearing stress, see Figure \ref{fig3}. The tangential electric field is discontinuous (as a result of the presence of transmembrane potential) producing vorticity and a behavior similar to the Kelvin-Helmholtz instability.  The shearing stresses decrease in time; once they vanish, the growth of the interface perturbation stops and it begins to relax back to the flat, unperturbed state. Increasing the applied voltage magnifies the amplitude of the perturbation in the shape of the membrane.

 \refeq{bc:perturbedcurrent}, and \refeq{bc:perturbedstress} along with base-state electric field $E_1^{(0)}(t)$ describe the evolution of the membrane perturbation.
This system can be written concisely as $\partial_t{\mathbf{g}}=\boldsymbol{\mathcal{C}}(t)\mathbf{g}$ where $\mathbf{g}=(h_q(t),A_q(t))^\text{T}$. In general, a numerical solution of this system  is required to investigate the stability of the membrane. Analytical results, however, can de derived if we  adopt the frozen coefficient approximation. In this approach, the matrix $\boldsymbol{\mathcal{C}}(t)$ is assumed constant allowing us to seek a solution of the form ${g}(x,z,t) ={g}(z)\exp[iqx+\omega t]$. The growth rates $\omega_{1,2}$ are found as eigenvalues of $\boldsymbol{\mathcal{C}}(t^*)$ for a fixed time $t^*$. Although this approach is only an approximation, the stability results are consistent with our numerical solution of the linear system of equations.

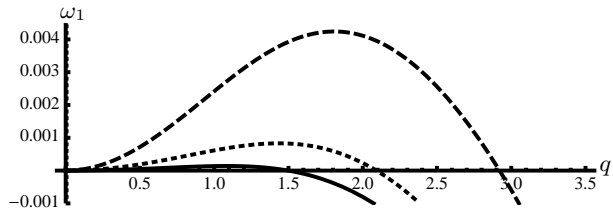
\begin{figure}[h]
\centerline{\includegraphics[height=1in]{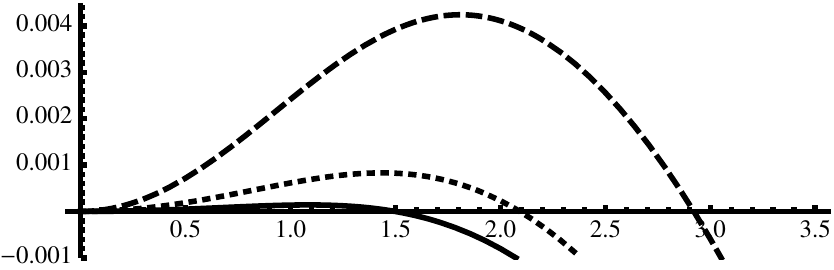}}
\begin{picture}(0,0)(0,0)
\put(112,26){$q$}
\put(-92,84){$\omega_1$}
\end{picture}
      \caption{\footnotesize   Time evolution of the dispersion  relation at fixed voltage. The dashed, dotted, and solid lines correspond to $t^*=0$, $t^* = 1$, and $t^* = 2$, respectively. The parameters are: $R=2$, $S=1$, $E= 1.4\, kV/cm$,  $\kappa = 10^{-19} \text{J}$ ,  $\Sigma = 10^{-9}\text{J/m}^2$, $\epsilon = 5.68\times10^{-10}\text{F/m}$, $\sigma_1 = 10^{-4} \text{F/m.s}$, and $C_m = 10^{-2} \text{F/m}^2$.}
      \label{fig4}
\end{figure}

In Figure~\ref{fig4} the dispersion relationship is shown for various frozen times at a fixed voltage of $V=0.14$. As time progresses,  the number of unstable wavenumbers decreases and eventually reaches zero.
For small $q$ the dispersion relationship seen in Figure \ref{fig4} can be approximated by
\begin{eqnarray}\begin{split}\label{expansion}
	\omega_1\sim \frac{\theta(t^*)}{12}q^2-\frac{1}{360}\bigg[\frac{30\Pi_1\zeta}{1+\lambda}+
	 \theta(t^*)\left( 6+5\delta(t^*) \right) \bigg]q^4
\end{split}\end{eqnarray}
%The effects of bending rigidity enter at $O(q^6)$. From the above expression, 
where $\theta(t^*) = \Pi_2(E_1^{(0)}(t^*))^2\mathcal{R}_1/(1+\lambda)$, $\mathcal{R}_1 = (R-1)(R^2-S)/(1+R)$ and $\delta(t^*) = \theta(t^*)(1+R)$.  The first term shows that the electric field acts as negative tension to destabilize the membrane; its magnitude diminishes with time and eventually the restoring bending and membrane tension forces take control. Setting $\omega_1$ to zero in \refeq{expansion} allows us to identify a cutoff wave number. The result confirms that instability can only occur if $(R-1)(R^2-S)>0$. Even in the absence of permittivity mismatch ($S=1$),  the instability occurs if the bulk fluid conductivities differ. If $S\neq 1$, then $(R-1)(R-\sqrt{S})$ must be positive to produce an instability for a nonzero electric field.

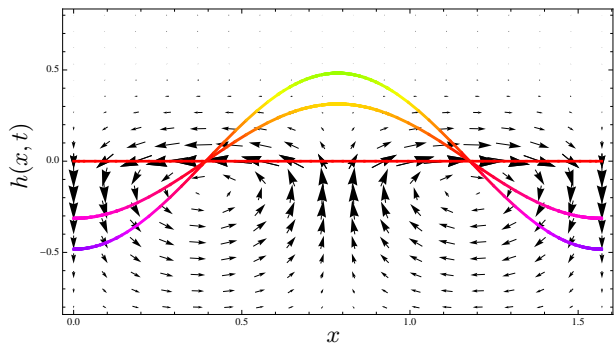
\begin{figure}[h]
\centerline{\includegraphics[width=3in]{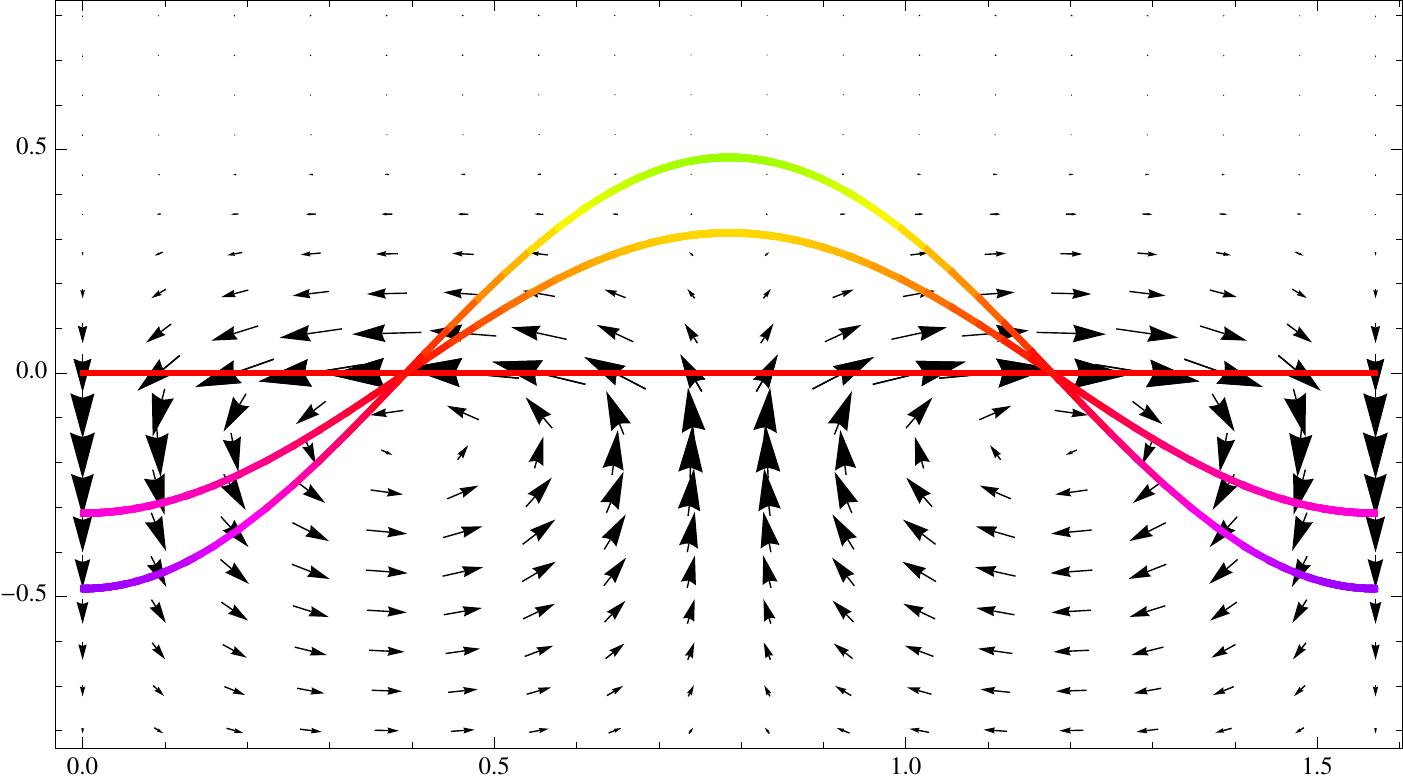}}
\begin{picture}(0,0)(0,0)
\put(0,5){$x$}
\put(-120,63){\rotatebox{90}{$h(x,t)$}}
\end{picture}
      \caption{\footnotesize  (Color online) Increasing shape undulation in response to a  perturbation in the transmembrane potential. The arrows show the fluid velocity field at $t=0$.  The color coding reflects the magnitude of lipid density where lighter shades indicate a decrease in the density (e.g the center region). Parameters as in Figure \ref{fig4}.}
      \label{fig5}
\end{figure}

We have also analyzed the electrohydrodynamics of a membrane modeled as two coupled monolayers  \cite{Seifert-Langer:1993, Schwalbe}.  
%In this model, a vanishing lipid density  is indicative of poration.coupled
In this case, the dynamics is described by four evolution equations for the interface height, electric field, and lipid densities of the two monolayers.
%The stability behavior is qualitatively similar. However, 
The ability to track the lipid density while the interface is deforming provides valuable additional information.
%, which allows to make conclusions regarding poration, i.e., vanishing lipid density. 
We find that during deformation  the density of the lipids on the top and bottom interface decreases in tandem, implying poration, see Figure \ref{fig5}. Full details can be found in \cite{Schwalbe:Thesis}.

In conclusion, we have shown that the combined action of  capacitive current through the membrane and mismatch of bulk conductivities destabilizes the membrane. This mechanism for creating  shape undulations may be relevant for budding and vesiculation during  electroformation.
%, where vesicles with relatively uniform size are produced.
 Moreover, considering a membrane modeled as two coupled monolayers, we find that the lipid density decreases in the regions of maximal deformation, which indicates poration in the context of the linearized model. A nonlinear investigation is needed to further clairify these results. 
 
JTS and MJM acknowledge financial support by NSF RTG grant DMS-0636574 and NSF grant DMS-0616468. PMV acknowledges partial financial support by NSF grant CBET-0846247.

\bibliographystyle{unsrt}
%\bibliography{refs}

\vspace{-.1cm}
\begin{thebibliography}{10}

\bibitem{Neumann-Sowers-Jordan:1989}
E.~Neumann, A.~E. Sowers, and C.~A. Jordan.
\newblock {\em Electroporation and electrofusion in cell biology}.
\newblock Plenum Press, New York, 1989.

\bibitem{Weaver:1996}
J.~C. Weaver and Y.~A. Chizmadzhev.
\newblock {\em Bioelectrochem.Bioenerg.}, 41:135--160, 1996.

\bibitem{Teissie:2005}
J.~Teissie, M.~Golzio, and M.~P. Rols.
\newblock {\em BBA}, 1724:270--280, 2005.

\bibitem{softmatter:2009}
R.~Dimova et al.
\newblock {\em Soft Matter}, 5:3201 -- 3212, 2009.

\bibitem{Needham-Hochmuth:1989}
D.~Needham and R.~M. Hochmuth.
\newblock {\em Biophys. J.}, 55:1001--1009, 1989.

\bibitem{Riske-Dimova:2005}
K.~A. Riske and R.~Dimova.
\newblock {\em Biophys. J.}, 88:1143--1155, 2005.

\bibitem{Melcher-Smith:1969}
J.~R. Melcher and C.V. Smith.
\newblock {\em Phys. Fluids}, 12:778--790, 1969.

\bibitem{Sens-Isambert:2002}
P.~Sens and H.~Isambert.
\newblock {\em Phys. Rev. Lett.}, 88:128102, 2002.

\bibitem{Ambjornsson:2007}
T.~Ambjornsson, M.~A. Lomholt, and P.~L. Hansen.
\newblock {\em Phys. Rev. E}, 75:051916, 2007.

\bibitem{Vlahovska-Dimova:2009}
P.~M. Vlahovska, R.~S. Gracia, S.~Aranda-Espinoza, and R.~Dimova.
\newblock {\em Biophys. J.}, 96:4789--4803, 2009.

\bibitem{Lacoste:2009}
D.~Lacoste, G.~I. Menon, M.~Z. Bazant, and J.~F. Joanny.
\newblock {\em Eur. Phys. J. E}, 28:243--264, 2009.

\bibitem{DeBruin-Krassowska:1999a}
K.A. DeBruin and W.~Krassowska.
\newblock {\em Biophys. J.}, 77:1213--1224, 1999.

\bibitem{Krassowska:2007}
W.~Krassowska and P.D. Filev.
\newblock {\em Biophys. J.}, 92:404--417, 2007.

\bibitem{Crowley:1973}
J.~M. Crowley.
\newblock {\em Biophys. J.}, 13:711--724, 1973.

\bibitem{Dimitrov:1984}
D.S. Dimitrov.
\newblock {\em J. Membrane Biology}, 78:53--60, 1984.

\bibitem{Isambert:1998}
H.~Isambert.
\newblock {\em Phys. Rev. Lett.}, 80:3404--3407, 1998.

\bibitem{Lacoste:2007}
D.~Lacoste, M.C. Lagomarsino, and J.F. Joanny.
\newblock {\em Europhys. Lett.}, 77:18006, 2007.

\bibitem{Lacoste:2010}
F.~Ziebert, M.~Z. Bazant, and D.~Lacoste.
\newblock {\em Phys. Rev. E}, 81:031912, 2010.

\bibitem{Saville:1997}
D.~A. Saville.
\newblock {\em Annu. Rev.Fluid Mech.}, 29:27--64, 1997.

\bibitem{Baygents-Saville:1988}
J.~C. Baygents and D.~A. Saville.
\newblock {\em Drops and bubbles: third international colloquium}, pages 7--17,
  1988.

\bibitem{Zholkovskij:2002}
E.~K. Zholkovskij, J.~H. Masilyah, and J.~Czarnecki.
\newblock {\em J. Fluid Mech.}, 472:1--27, 2002.

\bibitem{Seifert:1999}
U.~Seifert
\newblock {\em Eur. Phys. J. B}, 8:405--415, 1999

\bibitem{Seifert-Langer:1993}
U.~Seifert and S.A. Langer.
\newblock {\em Europhys. Lett.}, 23:71--76, 1993.

\bibitem{Schwalbe}
J.~Schwalbe, P.~M. Vlahovska, and M.~Miksis.
\newblock {\em J. Fluid Mech.}, 647:403--419, 2010.

\bibitem{Schwalbe:Thesis}
J.~Schwalbe.
\newblock{Ph.D. Thesis.} Northwestern University.

\end{thebibliography}

\end{document}